# Low temperature silicon epitaxy on hydrogen terminated Si(100) surfaces

Jeong-Young Ji and T.-C. Shen

Department of Physics, Utah State University, Logan, Utah 84322

Si deposition on H terminated Si(100)-2x1 and 3x1 surfaces at temperatures 300-530 K is studied by scanning tunneling microscopy. Hydrogen apparently hinders Si adatom diffusion and enhances surface roughening. The post-growth annealing effect is analyzed. Hydrogen is shown to remain on the growth front up to at least 10 ML. The dihydride units on the 3x1 surfaces further suppress the Si adatom diffusion and increase surface roughness.

68.55.-a, 68.35.Bs, 61.16.Ch

# 1. Introduction

The notion of epitaxial thickness was first raised by Eaglesham *et al.* by pointing out that Si deposition on Si(100) surfaces at temperatures between 323 and 573 K can be epitaxial up to a finite thickness, beyond which the epitaxy breaks down abruptly by a crystalline-amorphous transition.[1] The cause for the crystalline-amorphous transition, therefore, has been a subject of intensive investigations. Since hydrogen is the most abundant residual gas in any stainless steel chamber, it has been suspected to be responsible for the breakdown of Si epitaxy at low temperatures from the beginning. Wolff *et al.* first showed that molecular hydrogen, even with very low sticking coefficient on Si surfaces, can adversely affect the crystal structure of the grown Si film for growth temperatures below the H desorption temperature about 783 K.[2] They attributed the breakdown of epitaxy to a critical hydrogen coverage at the growth front which is attained by hydrogen segregation. More careful studies by introducing atomic hydrogen/deuterium during Si growth suggest that hydrogen limiting of Si diffusion is the primary cause for the breakdown of epitaxy.[3,4] The capability to overcome the diffusion barrier for adatoms may explain why the epitaxial thickness is sensitive to growth temperature in ambient hydrogen even below the hydrogen desorption temperature.

Hydrogen termination on Si(100) surface has two stable configurations: monohydride and dihydride, with a saturation H coverage of 1 and 2 monolayer (ML), respectively. The role of dihydride in the breakdown of epitaxy was first noted by Copel and Tromp[5]. Based on the fact that when the H coverage is greater than 1 ML, a sharp temperature increase is required for epitaxy, they suggested that dihydride may disrupt the dimer structure of Si(100) or act as a nucleation site which leads to surface roughening. The adverse effect of dihydride on Si homoepitaxy can be further elucidated from the observation that on HF dipped samples, epitaxy requires a growth temperature > 643 K.[6,7] This can be explained by the fact that at ~680 K most of the dihydride units on HF dipped surfaces are converted to monohydride by $H_2$ desortion[8-10], and the monohydride surface is more apt for epitaxy (see our results in Sec. 3). Indeed, theoretical works reveal that the activation energy for Si adatom diffusion on a dihydride surface is dramatically higher than either the bare or the monohydride surface.[11,12] The overall picture suggests that both monohydride and dihydride on the Si surface hinder the Si adatom diffusion, which disrupts the layer-by-layer growth on the bare surface and builds up surface roughness which eventually leads to crystalline-amorphous transition.[13]

Most of the aforementioned studies of the crystallinity of a thin film were conducted by transmission electron microscopy (TEM), which is an excellent tool to characterize the film quality in cross section, but not easy to extract roughness information in a plane view. To correlate the surface roughness with epitaxial breakdown, Karpenko *et al.* conducted a reflection high energy electron diffraction (RHEED) study on Si film growth at 473-548 K on bare Si(100) surfaces and found that the surface roughness increases linearly with the film thickness, at least initially, at a roughening rate that decreases with increasing growth temperature.[13] A scanning tunneling microscopy (STM) study on the surface morphology of Si deposition onto sub-0.1 ML H-covered Si(100) surfaces at substrate temperatures 300-500 K, revealed a significant increase of Si island density compared to the bare surface, and the diffusion barrier, twice as large as those on the bare surface.[14] However, above 550 K, the authors in Ref. 14 concluded that the H influence disappears. In Ref. 5 Copel and Tromp observed that 20 ML Si film can be epitaxial at a growth temperature < 410 K when the H coverage of the substrate is <1 ML, but requires a sharp temperature increase to 470 K when the H coverage is above 1 ML, based on the result of medium energy ion scattering (MEIS).

Part of the uncertainty comes from the initial surface preparation. Dipping in HF does not create a uniform dihydride surface.[15] Si deposition in a hydrogen-rich ambience intending to emulate chemical vapor deposition introduces further complications such as H adsorption and abstraction into the equation. However, uniform monohydride surfaces with 2x1 reconstruction and ordered monohydride-dihydride mixed surfaces with 3x1 reconstruction can be readily prepared in ultrahigh vacuum (UHV).[16] We think a real-space atomic scale investigation on the initial stage of the surface roughening during Si homoepitaxy on well-controlled H terminated Si(100) surfaces will help to clarify the H role in Si homoepitaxy. In addition, since the epitaxial thickness of a Si film on H terminated Si(100) surfaces is much less than that on a bare surface, this study may shed some light on the surface morphology of crystalline-amorphous transition.

In this paper, we report our STM study of the surface morphology of 1 to 20 ML Si deposition onto the UHV prepared H terminated Si(100)-2x1 and 3x1 surfaces at temperatures between 300 and 530 K. The surface roughening enhanced by silicon monohydride and dihydride is clearly visible. The underlying lattice structure is still epitaxial despite apparent surface roughness. The effect of post-deposition annealing on the surface is examined. Finally, we use STM induced electron stimulated desorption to confirm a complete H segregation after Si deposition up to 10 ML.

## 2. Experimental procedures

Samples were cut from either p- or n-type Si(100) wafers with resistivity 0.1 Ω-cm. After a standard RCA[17] cleaning, the samples were introduced into the UHV chamber with base pressure ~$1\times10^{-10}$ Torr and degassed at 920 K by resistive heating for 8 h. Clean 2x1 surfaces were obtained by flashing the samples to 1500 K for 1 min while keeping the chamber pressure <$1\times10^{-9}$ Torr. The sample temperature was measured by a C-type thermocouple attached to the back side of the sample for low temperatures and by a pyrometer for $T > 873$ K. The thermocouple was further calibrated at the room temperature and the H-desorption temperature (783 K). H-termination was achieved by positioning a freshly annealed Si sample 6 cm from a hot W-filament in the UHV chamber back filled with $1\times10^{-6}$ Torr $H_2$ for 5 min at ~620 K to prepare 2x1 monohydride surfaces and at ~400 K to prepare 3x1 surfaces. Si deposition was performed by a home-made e-beam evaporator. The deposition was monitored by a crystal microbalance and the rate was 0.5-0.7 Å/min. The ambient pressure during Si deposition was < $4\times10^{-10}$ Torr. All the STM images presented in this work are filled state images, taken at a sample bias of -2 V and tunneling current 0.1 nA. Each ML deposition corresponds to 0.14 nm in thickness.

# 3. Results

**3.1 Si deposition on H/Si(100)-2x1 at $T \leq 470$ K**

Short rows and small clusters were observed after 1 ML Si deposition at room temperature (RT) on monohydride Si(100) surfaces as shown in Fig.1, which is consistent with the submonolayer deposition result of Ref. 14. Apparently the kinetic energy of the ad-atoms can not overcome the diffusion barrier to form any long range ordered structures. Even at 470 K, we find the surfaces are clustered with three-dimensional (3D) islands while the underlying 2x1 surface structure is still visible (Fig. 2). The average diameter of the 3D islands after 2 ML and 10 ML Si deposition is ~1.5 nm and 3 nm, respectively while the surface roughness ~0.4 nm and 0.9 nm, respectively. These numbers suggest that growth of the 3D islands is preferred to nucleation of new islands on the H terminated surfaces. Larger scale images in Fig. 3 show that the original terraces are still visible after 10 ML Si deposition at 470 K. These results demonstrate that even though both the bare and monohydride Si(100)-2x1 surfaces have the same reconstruction symmetry, the interaction of the Si adatoms with the substrate is dramatically different and the diffusion of the adatoms is much hampered on the monohydride surface.

**3.2 Si deposition on H/Si(100)-2x1 at *T*~530 K**

Depositing Si at higher substrate temperatures certainly helps the adatom diffusion and improves the long range order. Fig. 4(a) shows the result of 2.5 ML deposition on a bare Si(100) surface at 490 K. The large anisotropic two-dimensional (2D) islands are in great contrast to the much smaller islands in Fig. 4(b) where the same amount of Si is deposited on a monohydride surface at 530 K. The film is clearly epitaxial, but the presence of smaller epitaxial islands is consistent with the scenario of higher diffusion barrier on the monohydride surfaces. The overall 3 ML surface roughness is, however, unchanged for both bare and monohydride surfaces.

Continuous deposition of Si at 530 K seems to be able to rearrange the domain boundaries and fill in vacancies in lower levels to preserve the epitaxial registration. Fig. 5(a) shows the surface after 10 ML deposition. The 3 ML surface roughness suggests that earlier defects have been mended with further Si deposition. The surface after 20 ML deposition at 530 K is shown in Fig. 5(b). The original terraces are no longer visible due to the 5 ML roughness. However, the mutually perpendicular orientation between islands in different terraces indicates local epitaxy albeit reduced island sizes. These images suggest that the thin film growth involves more than the surface layer atoms. With sufficient substrate temperature, it is energetically favorable for atoms of a few layers deep to rearrange themselves continuously to achieve epitaxy.

**3.3 Post-deposition annealing**

Fig. 6(a) shows the surface after 5 ML deposition at 530 K. The ratio of the areas on the first, second and third layer is ~20 %, 73 % and 98 %, respectively. The result of 5 min annealing at 680 K changes the ratio of the layers to ~14 %, 82 % and 98 %, respectively. Comparing the annealed 5 ML deposition in Fig. 6(b) with unannealed 3 ML in 4(b), one can conclude that the effect of the annealing is to restore the crystallinity of the deposited film by moving the top layer atoms downward to fill in vacancies in the lower layer. Even at 680 K the diffusion of the adatoms is still limited, so the step morphology shown in Fig. 7(a) is not much different from the pre-deposited surface (not shown). However, annealing 5 min at 880 K allows the adatoms to diffuse to the step edge and the surface becomes essentially flat with a few single layer 2D islands (Fig. 7b).

**3.4 Hydrogen segregation**

H segregation during growth from a H-terminated substrate was first reported by Copel and Tromp[5]. They estimated H concentration in the film is <1 % at growth temperature > 393 K. With atomic resolution we can check the above claim by STM. After 5 ML deposition on a monohydride Si(100)-2x1 surface at 530 K followed by annealing 5 min at 680 K, the central region of Fig. 8 is irradiated with electrons from the STM tip at a sample bias of 7 V and field emission current of 0.1 nA. The H atoms in this region are thus removed by the electron stimulated desorption[18]. The Si dangling bonds appear brighter in contrast to the rest of the surface where the Si dangling bonds are terminated by H. Line A resides in the H-terminated region and line B is in the H-desorbed region. It is clear from the two linescans in Fig. 8 that the apparent heights of the dimer rows in both regions are in unit of single layer distance (0.14 nm). Note that a bare Si dangling bond and a bare Si dimer would appear 0.12 nm and 0.16 nm higher, respectively, than a Si monohydride dimer in the filled state STM image. This result indicates that at least during the initial stage of Si deposition on monohydride surface, very few H atoms are desorbed. The dimer strings on the top layer are completely H terminated just as those in the lower layers.

**3.5 Si deposition on H/Si(100)-3x1 at 530 K**

To investigate if the silicon dihydride plays a significant role in the surface roughening during Si homoepitaxy, we prepared H/Si(100)-3x1 surfaces which consist of alternating monohydride and dihydride rows.  We found that both 3 ML and 10 ML Si deposition on the 3x1 surface at 530 K result in short dimer rows and 3D clusters. The underlying dimer rows and terraces are still visible. The 10 ML deposited surface is depicted in Fig. 9 which is significantly different from the corresponding deposition on the 2x1 surface at 530 K [Fig. 5 (a)]. The surface roughness is ~1 nm compared to 0.4 nm on 2x1. From this result, we would expect that a film grown on a disordered dihydride surface at 473 K should be totally amorphous as observed in Ref. 19.

## 4. Discussion and conclusions

In order to demonstrate unambiguously the diffusion barrier increase due to surface silicon hydride and H segregation during silicon growth, we have conducted Si deposition on well ordered H/Si(100)-2x1 and 3x1 surfaces. Such surfaces should exclude any potential effects related to surface disorder or multiple hydride caused by wet-chemical preparation. Unlike many studies in the literature, we do not deliberately introduce atomic hydrogen during Si deposition in order to minimize H adsorption and abstraction which introduces further complications.

Comparing Fig. 4(a) and 4(b), it is clear that adatoms diffusion on the two surfaces with identical symmetry are quite different at 530 K. There are smaller, less anisotropic islands on the monohydride surface than on the bare Si(100) surface, which is consistent with the theoretical conclusion that the activation energies on the monohydride surface are 1.5 and 1.7 eV along the direction parallel and perpendicular to the dimer rows, respectively[12], as opposed to the corresponding 0.6 and 1.0eV on the bare surface[20].

We find that Si overgrowth on the monohydride Si(100) is amorphous at least below 470 K implying that the adatoms are unable to overcome the diffusion barrier effectively. Raising the growth temperature to 530 K allows epitaxial 2D islands to form albeit small in size. However, even the surface roughening increases with growth, the roughness is far less than the film thickness implying multi-layer rearrangement during deposition. Post-growth annealing moves the top layer atoms downward to fill in vacancies and coalescence the 2D islands to result in a surface similar to that of a thinner, smoother film.

In order to have epitaxial growth, the surface Si-Si dimer bonds must be broken. The 3x1 reconstructed surface actually has only half of the dimer bonds but has 1/3 ML more H coverage than the 2x1 reconstructed monohydride surface. Apparently, the diffusion barrier created by dihydride plays a more significant role in the surface roughening process than the energy cost to break dimer bonds as can be seen clearly by comparing Fig. 5(a) with Fig. 9. The much higher activation energy, 2.2 eV, for Si ad-atom diffusion on 3x1 surfaces[11], is certainly consistent with our results.

We have presented an atomic scale evidence of the nearly complete H segregation during Si deposition up to at least 10 ML on the 2x1 monohydride surface. The intriguing interaction of the incident Si adatom and the substrate Si-H unit was first considered by Murty and Atwater[19]. Based on the results of molecular dynamics simulation, they proposed that the incident Si atom can either bond to the subsurface level and hence the whole SiH unit can be segregated or an exchange mechanism allows the H atom to transfer from a substrate Si atom to the incident Si atom.[19] Recently a no barrier pathway[11] was discovered which allows the adatom atom to bond directly to a dimer atom and the H atom, originally bonded to the dimer atom, forms a bond with the adatom . The SiH unit then can diffuse either along or perpendicular to the dimer row with a barrier 0.7 eV and 1.0 eV, respectively.

In summary, H atoms on the Si(100) surface play an important role in the Si homoepitaxy. The SiH and $SiH_2$ unit can block Si adatom diffusion to increase surface roughness, but they also exchange with the incident Si adatom to keep H as a surfactant on the growth front. Recently, Bratland *et al.* showed the evolution of surface morphology during low-temperature Ge homoepitaxy by atomic force microscopy.[21] The surface has a roughness of 4 nm after a 7 nm deposition at 428 K. Further deposition leads to the formation of self-organized mounds as the precursor of {111} facets and cusps before epitaxy breaks down. Our study fits in at the very early stage of the evolution when the surface roughness is only a few monolayers.

## 5. Acknowledgements

The experiments reported here would not be possible without the silicon e-beam evaporator made by Toby Barrus. We are also grateful to Y. C. Chang and G. Qian for sharing with us their theoretical results before publication. This work was supported in part by the National Security Agency (NSA) and Advanced Research and Development Activity (ARDA) under Army Research Office (ARO) contract number DAAD19-00-1-0407, the DARPA-QuIST program under contract number DAAD 19-01-1-0324 and by the National Science Foundation under grant number DMR-9875129.

**Figure captions**

**Fig.1** 1 ML Si deposition at room temperature on a H/Si(100)-2x1 surface.

**Fig.2** (a) 2 ML and (b) 10 ML Si deposition on a H/Si(100)-2x1 at 470 K.

**Fig.3** (a) 2 ML and (b) 10 ML Si deposition on a H/Si(100)-2x1 at 470 K.

**Fig.4** (a) 2.5 ML Si deposition on a bare Si(100)-2x1 surface at 490 K followed by H dosing to form a 3x1 surface. (b) 3 ML Si deposition on a H/Si(100)-2x1 at 530 K.

**Fig.5** (a) 10 ML (b) 20 ML deposition of Si on H/Si(100)-2x1 at 530 K.

**Fig.6** (a) 5 ML deposition of Si on H/Si(100)-2x1 at 530 K (b) After a 5 min annealing at 680 K.

**Fig.7** 5 ML deposition of Si on H/Si(100)-2x1 followed by a (a) 5 min annealing at 680 K (b) 3 min annealing at 880 K. The bright protrusions are Si dangling bonds.

**Fig.8** The same surface as Fig. 7(a) after the center region being irradiated by 7V electrons from STM to desorb H. The bare Si dangling bonds appear brighter than the rest of the surface. The three doted lines in the linescan A and B indicate the three atomic layers with 0.14 nm apart.

**Fig.9** 10 ML deposition of Si on H/Si(100)-3x1 at 530 K. The inset is a close-up image of 29x29 nm$^2$.

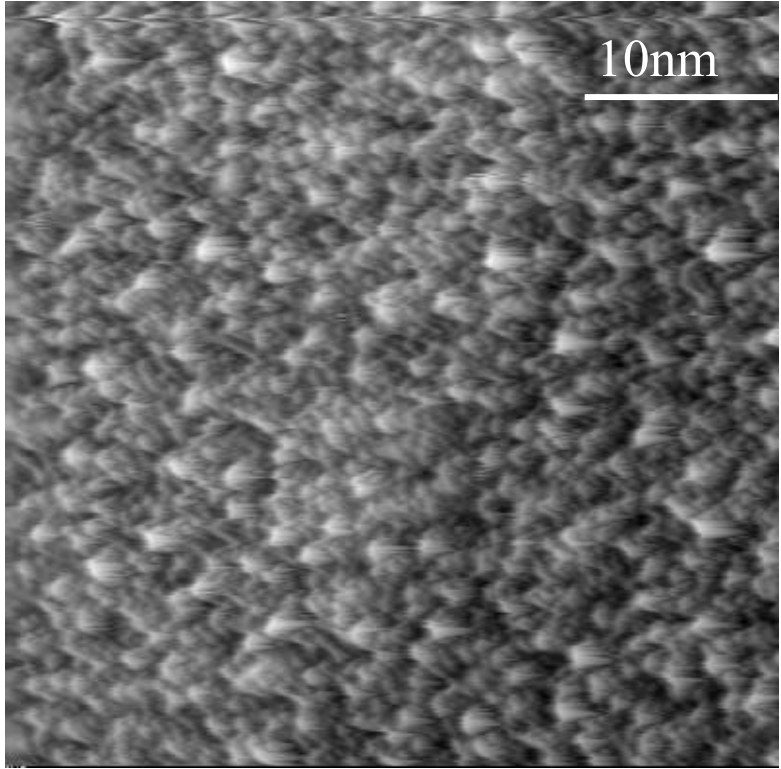

Fig. 1

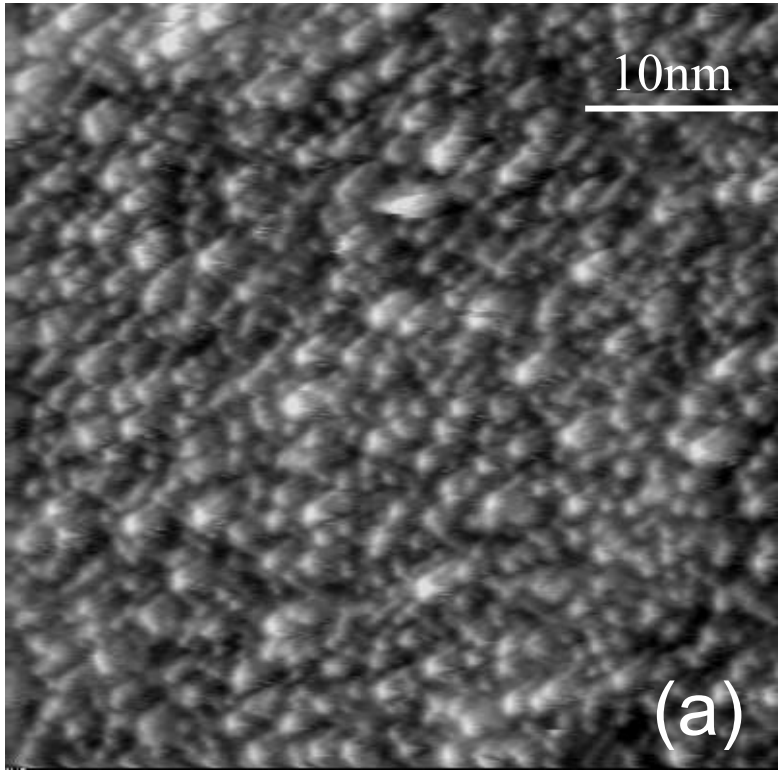
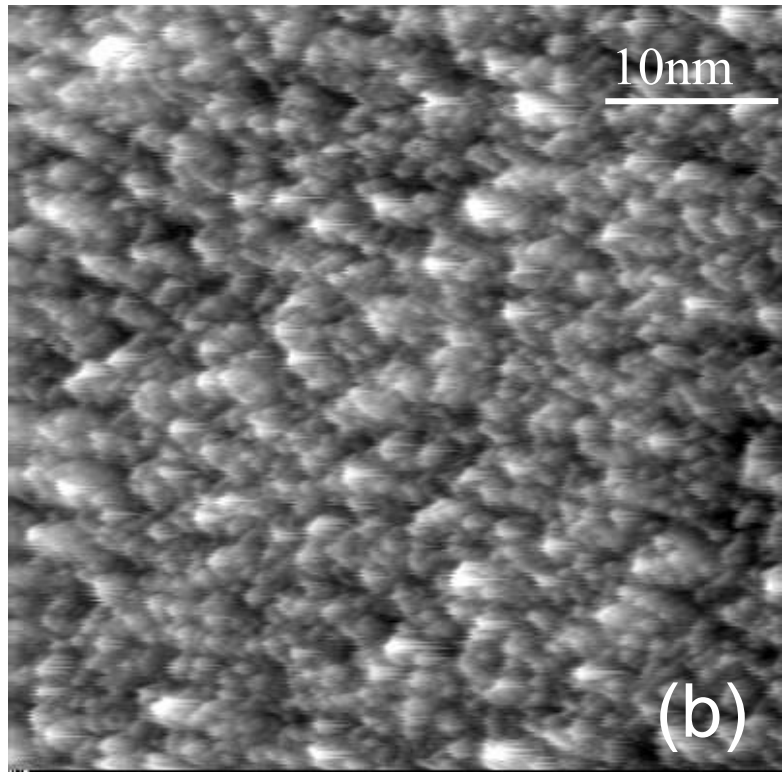

Fig. 2

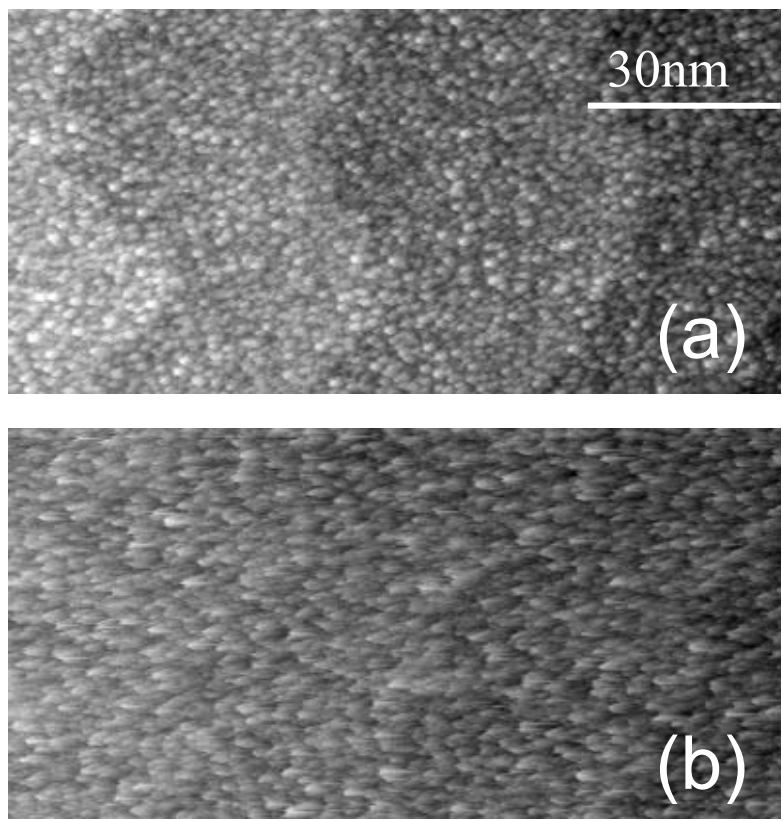

Fig. 3

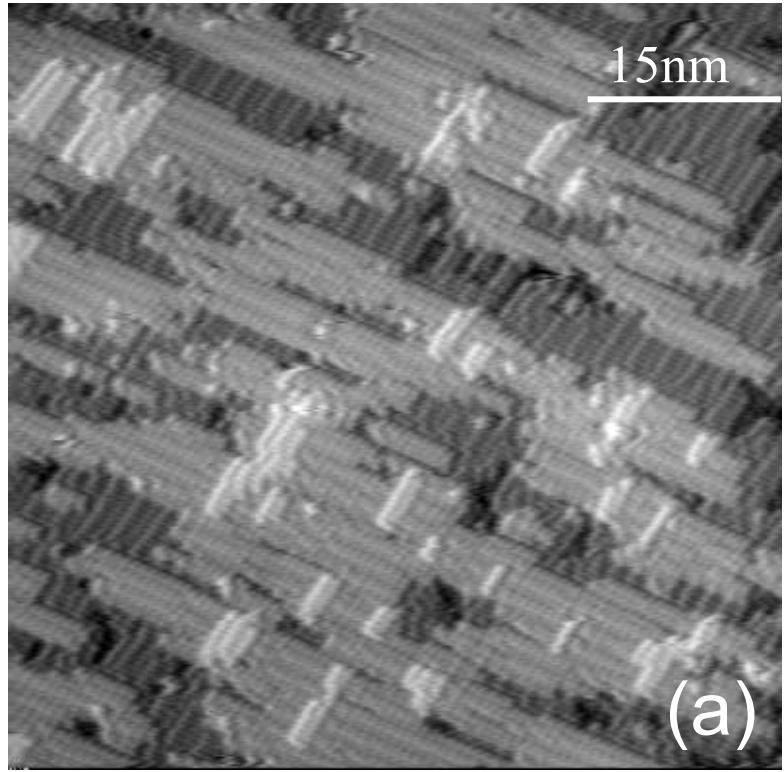

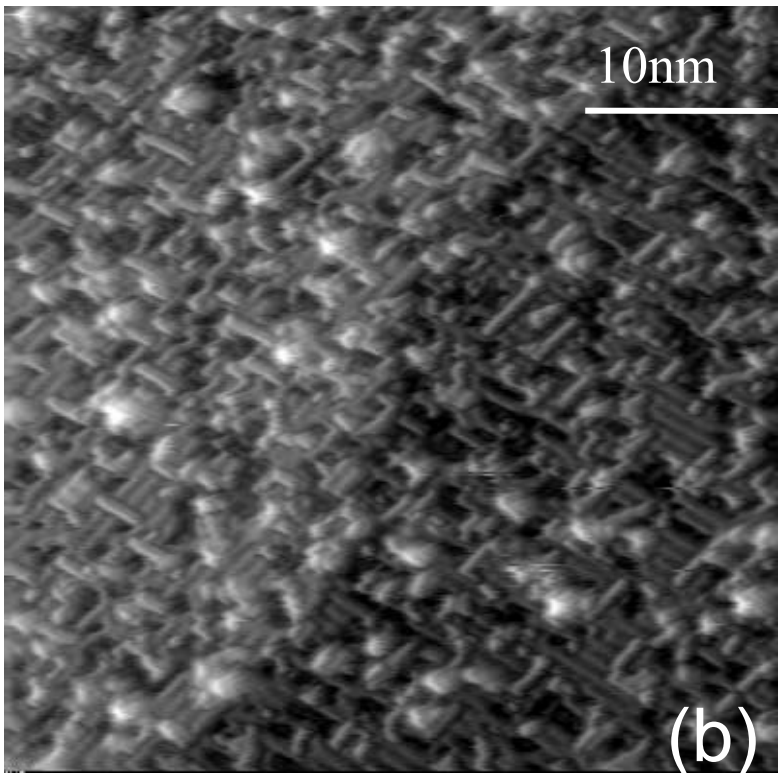

Fig. 4

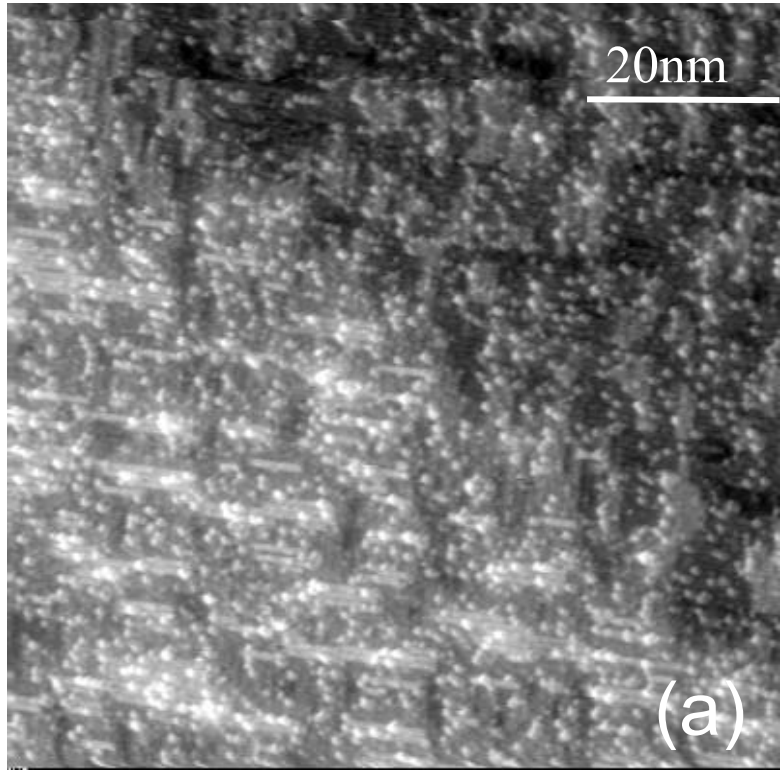

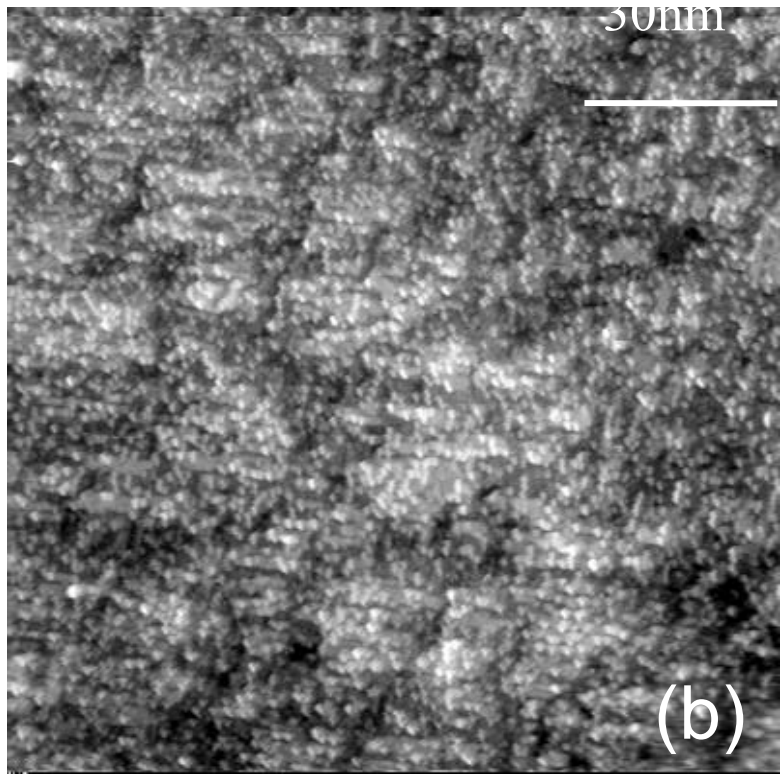

Fig. 5

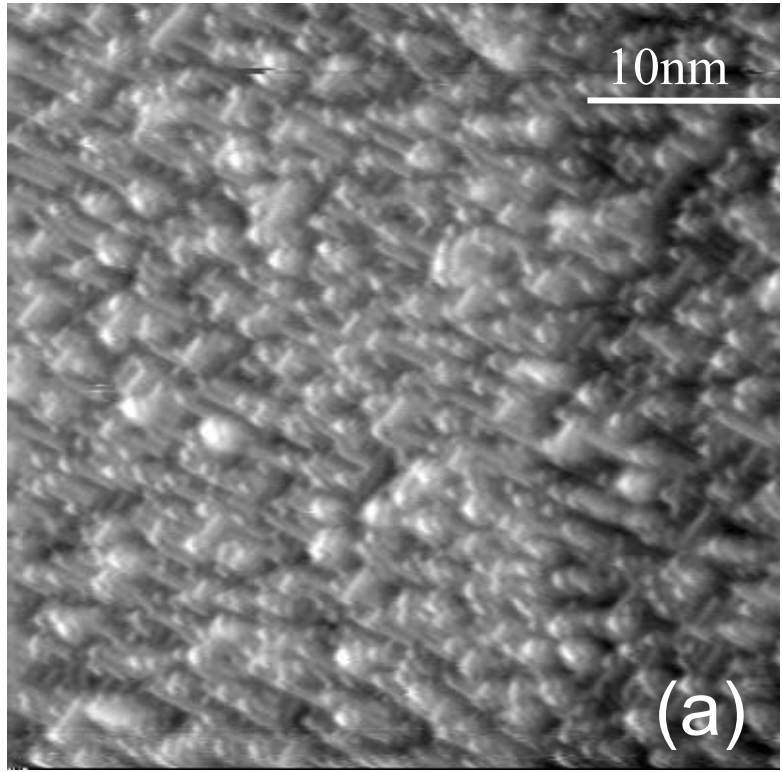

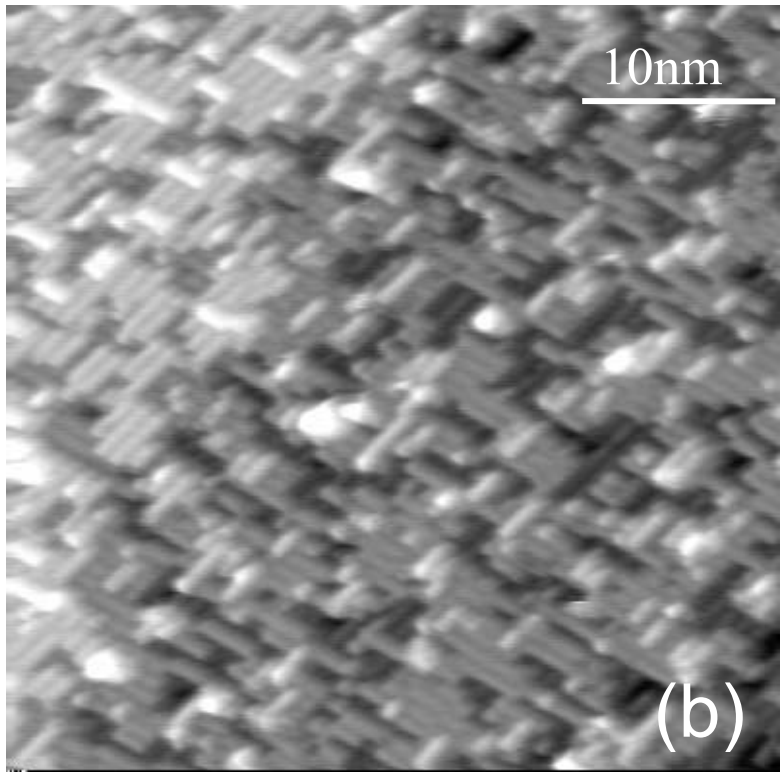

Fig. 6

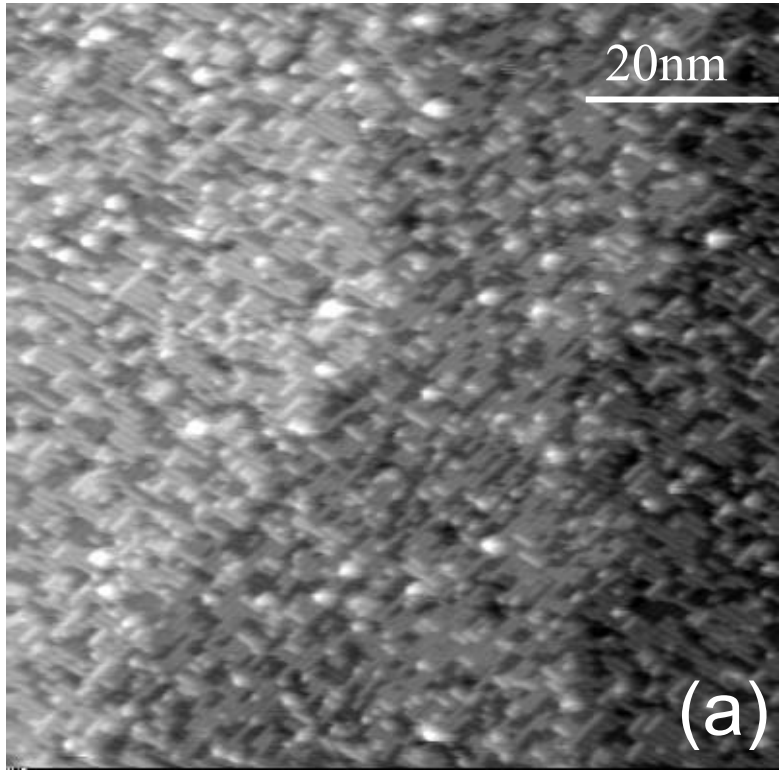

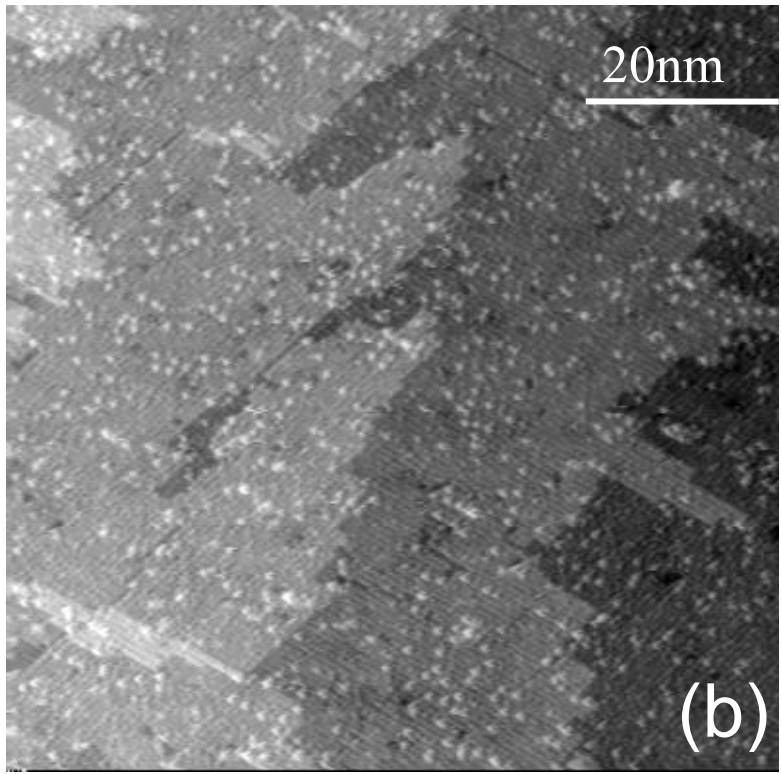

Fig. 7

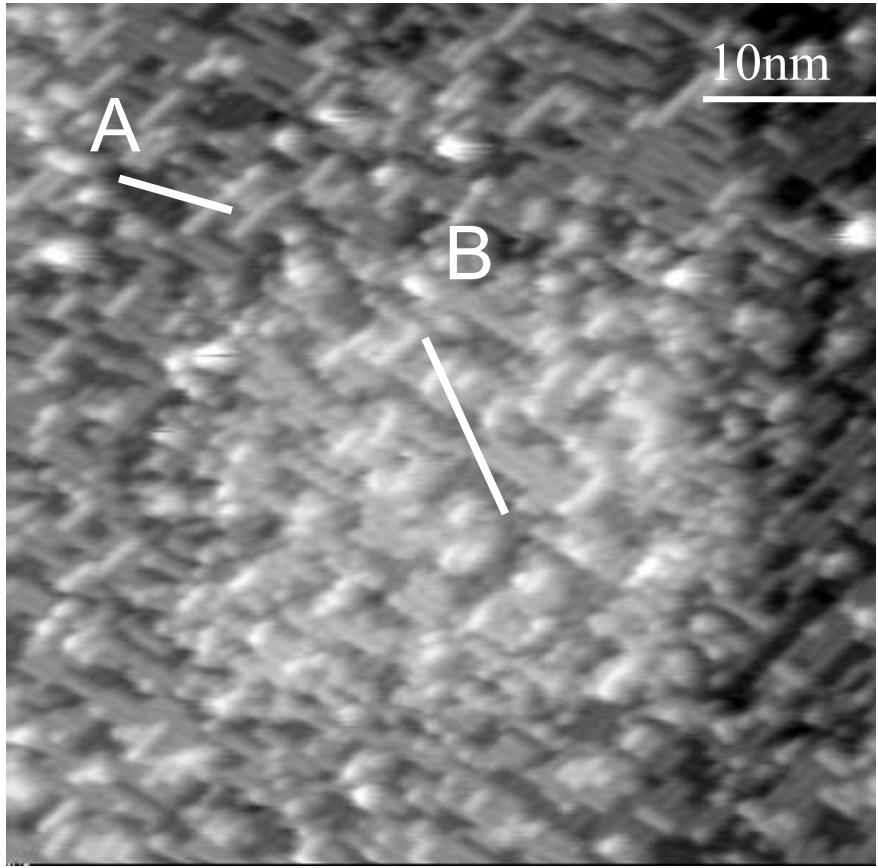

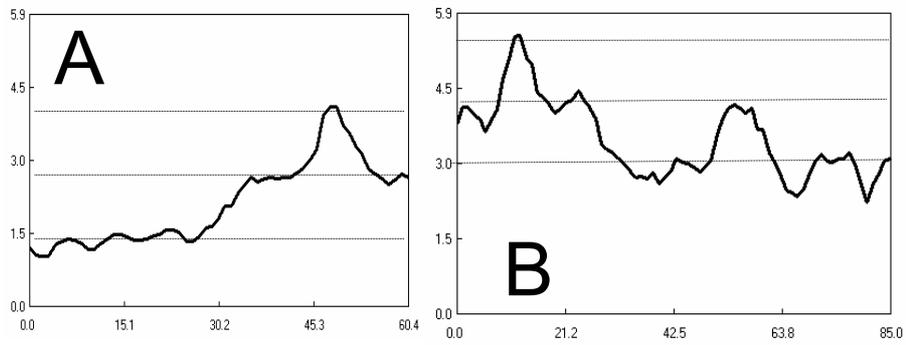

Fig. 8

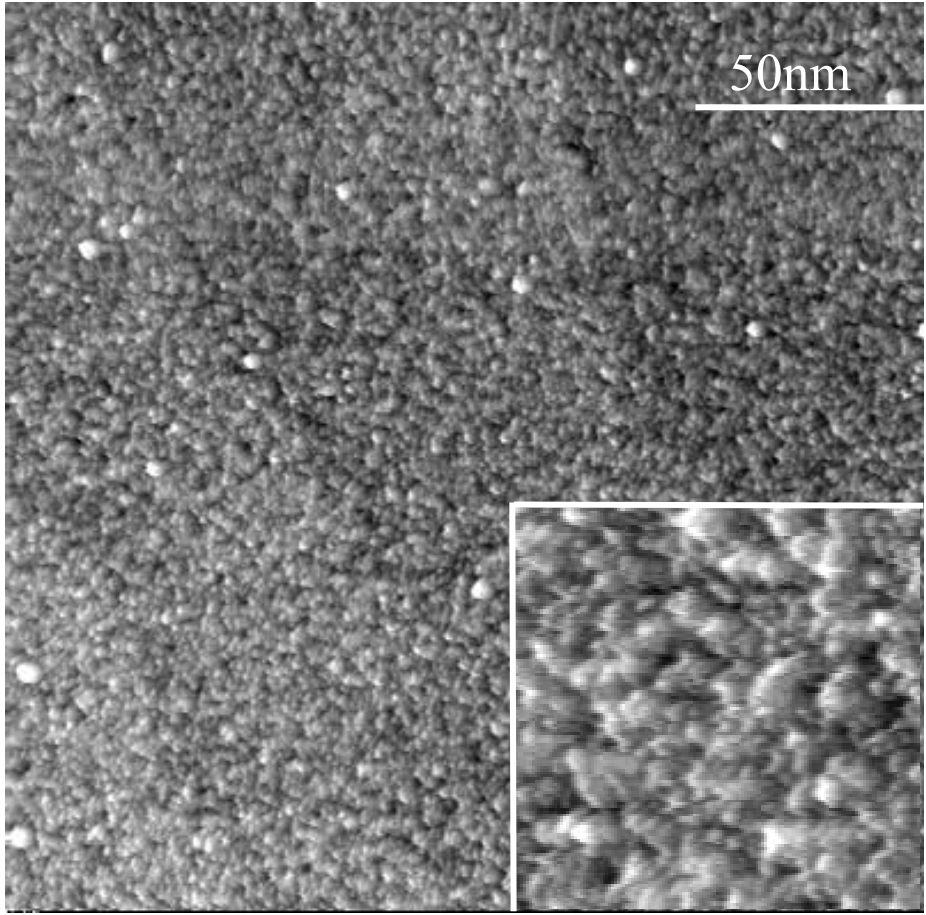

Fig. 9